\documentclass[a4paper]{nature}
\usepackage[utf8]{inputenc}
\usepackage[english]{babel}
\usepackage[]{graphicx}
\usepackage{amsmath}
\usepackage{amssymb}
\usepackage{mathrsfs}
\usepackage{textcomp}
\usepackage{tabularx}
\usepackage{color}
\usepackage{array}
\usepackage{float}
\usepackage{braket}
\usepackage{mathtools}

\definecolor{darkblue}{rgb}{0.1,0.2,0.6}
\definecolor{darkred}{rgb}{0.8,0.1,0.2}
\usepackage[colorlinks,citecolor=darkblue,linkcolor=darkred,urlcolor=darkblue]{hyperref}

\makeatletter
\let\saved@includegraphics\includegraphics
\AtBeginDocument{\let\includegraphics\saved@includegraphics}
\renewenvironment*{figure}{\@float{figure}}{\end@float}
\makeatother

\title{\begin{center}
		Spin Cross-Correlation Experiments\\
		in an Electron Entangler
	\end{center}}

\author{Arunav Bordoloi$^{1,^{*}}$, Valentina Zannier$^{2}$, Lucia Sorba$^{2}$, Christian Sch\"onenberger$^{1,3}$ \& Andreas Baumgartner$^{1,3,^{*}}$}

\begin{document}

\maketitle
\thispagestyle{empty}

\begin{affiliations}
	\item Department of Physics, University of Basel, Klingelbergstrasse 82, CH-4056 Basel, Switzerland
	\item NEST, Istituto Nanoscienze-CNR and Scuola Normale Superiore, Piazza San Silvestro 12, I-56127 Pisa, Italy
	\item Swiss Nanoscience Institute, University of Basel, Klingelbergstrasse 82, CH-4056, Basel, Switzerland
	\item[$^{*}$] Corresponding authors: arunav.bordoloi@unibas.ch, andreas.baumgartner@unibas.ch
\end{affiliations}

\begin{abstract}
	
Correlations are fundamental in describing many body systems - not only in natural sciences. However, in experiments, correlations are notoriously difficult to assess on the microscopic scale, especially for electron spins. Here, we demonstrate a direct measurement of the spin cross-correlations between the currents of a Cooper pair splitter, an electronic device that emits electrons originating from Cooper pairs in a superconductor. While it is firmly established theoretically that these electron pairs form maximally spin-entangled singlet states with opposite spin projections, no spin correlation experiments have been demonstrated so far. We use ferromagnetic sidegates, compatible with superconducting electronic structures, to individually spin polarize the transmissions of two quantum dots fabricated in the two electronic paths, which act as tunable spin filters. The signals are detected in standard transport and in highly sensitive transconductance experiments. We find that the spin-cross correlation is negative, compatible with spin singlet emission, and deviates from the ideal value mostly due to a finite overlap of the Zeeman split quantum dot states. Our results demonstrate a new route to perform spin auto- and cross correlation experiments in nanometer scaled electronic devices, especially suitable for those relying on magnetic field sensitive superconducting elements, like unconventional, triplet or topologically non-trivial superconductors, or to perform Bell tests with massive particles, like electrons.         	

\end{abstract}

\maketitle

\section*{Introduction}

Correlations are essential in almost all scientific fields and usually describe a relation between two (or more) observed variables. In modern physics, correlations are often used to characterize exotic quantum states,\cite{Sachdev2008,Broholm2020} or multi-qubit states in prospective quantum computers.\cite{Chatterjee2021} Spin correlations are especially interesting, since many thermodynamic phases in condensed matter are related to the electron spin, for example various magnetic phases.\cite{Fert2017,Coronado2019,Kurebayashi2022} However, to directly correlate one electron spin to another, i.e. to measure the so called spin cross-correlation, is still uncharted territory.

One of the most prominent examples of correlated electrons are superconductors, in which a large number of electrons forgo their individual fermionic nature and lower the ground state energy by forming a collective of bosonic electron pairs called Cooper pairs.\cite{Bardeen1957} As a consequence, only an even number of electrons can be added or removed at a time at low energies, with the electron pairs forming spin singlet states. The process of two such electrons tunneling into a normal metal is the well known Andreev reflection.\cite{Beenakker1997} If they are allowed to tunnel into two separate normal contacts,  

\begin{figure}[H]
	\centering
	\includegraphics[width=\columnwidth]{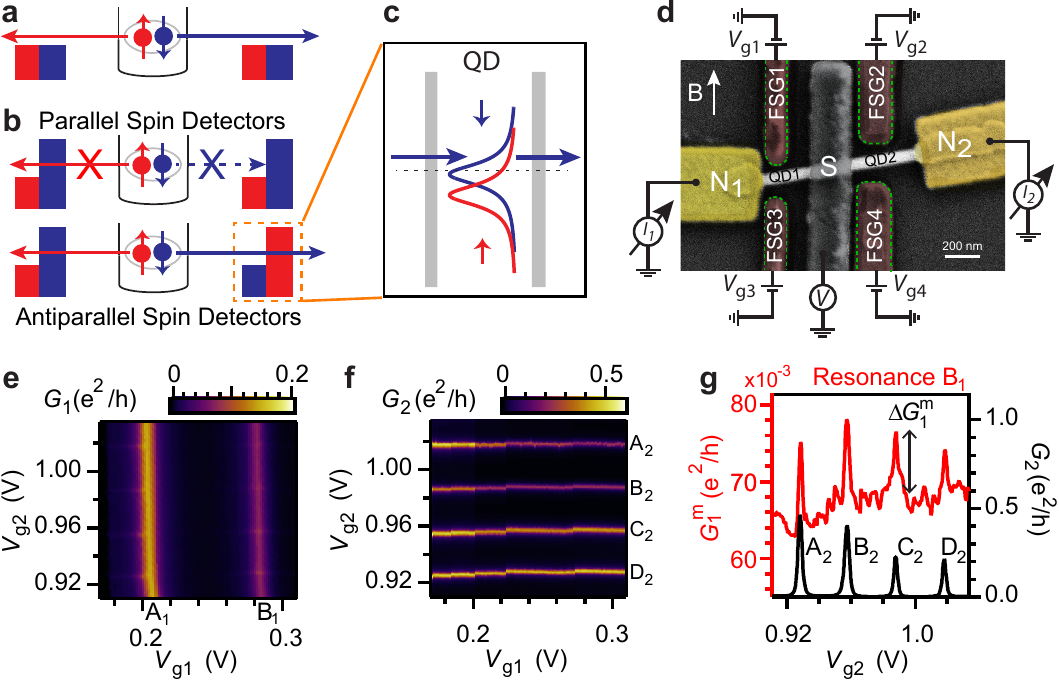}
	\caption{\textbf{Principles of spin and charge correlation CPS experiments.} \textbf{a} Illustration of the Cooper pair splitting process. \textbf{b} Schematic of spin correlation experiments with spin filters to the left and right. \textbf{c} Implementation of a spin detector by a Zeeman split quantum dot. \textbf{d} False color scanning electron microscopy image of the presented CPS device. \textbf{e,f} Differential conductances $G_{1}$ and $G_{2}$, respectively, measured simultaneously as a function of $V_{\textrm{g1}}$ and $V_{\textrm{g2}}$ at a source-drain bias voltage $V_{\textrm{dc}} = 0$. \textbf{g} Conductance maximum $G^{\textrm{m}}_{1}$ of QD1 resonance $\textrm{B}_{1}$ as a function of gate voltage $V_{\textrm{g2}}$, showing peaks when $G_{2}$ is tuned across the Coulomb blockade resonances A$_{2}$-D$_{2}$ of QD2.}
	\label{fig:Fig_1}
\end{figure}

\noindent an additional transport mechanism called crossed Andreev reflection (CAR)\cite{Russo2005,CaddenZimansky2009,Kleine2009} becomes available. This process can even become dominating if Coulomb repulsion on small semiconductor islands called quantum dots (QDs) enforce the separation of the charges, which is often called Cooper pair splitting (CPS), and illustrated in figure~\ref{fig:Fig_1}a. Such a device is expected to form a highly efficient source of spatially separated, maximally spin entangled electrons,\cite{Recher2001,Lesovik2001,Samuelsson2003} and, more fundamentally, allow for a Bell test with massive particles.\cite{Kawabata2001,Klobus2014} The charge correlations in a CPS device, i.e. the simultaneous emission of two electrons, was heavily investigated in recent years, both experimentally,\cite{Hofstetter2009,Herrmann2010,Das2012,Schindele2012,Fueloep2014,Tan2015,Lee2017,Ranni2021,Pandey2021} and theoretically,\cite{Recher2001,Futterer2009,Trocha2015,Trocha2018,Tam2021} stressing its fundamental and practical importance.\cite{Benjamin2006,Braunecker2013,Klobus2014,Busz2017} However, the experimental detection of the corresponding spin states and their correlations have proven to be very challenging,\cite{Beckmann2004,Scheruebl2014,Hels2016} especially since most proposed schemes rely on ferromagnetic contacts competing with superconductivity.

In this work, we directly measure the cross-correlation between the spin currents emitted from a superconductor (S) into two adjacent QDs in a Cooper pair splitter. The basic idea is illustrated in figure~\ref{fig:Fig_1}b. If both contacts to S act as spin filters, i.e. with a spin dependent single electron transmission, the probability for a Cooper pair in the singlet state $\ket{\rm cps}=\frac{1}{\sqrt{2}}\left(\ket{\uparrow \downarrow}-\ket{\downarrow \uparrow}\right)$ to be split into the two contacts depends on the relative orientation of the spin filters: for fully polarized spin filters oriented in parallel, we expect the probability for Cooper pair splitting to be fully suppressed, while for the anti-parallel orientation CPS is allowed (figure~\ref{fig:Fig_1}b). We use the spin correlation operator $\hat{C}=\sum_{\sigma_1,\sigma_2} s_1 s_2 \ket{\sigma_1\sigma_2}\bra{\sigma_1\sigma_2}$ with $\sigma_{j} \in \left\{ \uparrow , \downarrow \right\}$, $s_j=+1(-1)$ for $\sigma_j=\uparrow (\downarrow)$, i.e. with a positive value for parallel spins and a negative value for anti-parallel spins, where $j$ labels the QD. For the fully anti-correlated Cooper pair states, the expectation value for $\hat{C}$ should reach $\langle \hat{C}\rangle_{\rm cps} = -1$. All other possible transport processes\cite{Schindele2012,Fueloep2014,Ranni2022} are discussed in extended data table 1 and result in correlation values of either zero for processes involving only one QD, or positive values for higher order two-dot processes. In direct transport and transconductance experiments, we demonstrate a clear negative spin cross-correlation reaching values of $\langle \hat{C} \rangle_{\textrm{exp}} \approx -0.4$, limited by the polarization of the spin filters. Such devices provide a direct route towards spin correlation experiments at the nanometer scale in modern quantum-electronic devices suitable for various applications, such as the detection of triplet and topological superconductivity,\cite{Bergeret2005,Jeon2021} the investigation of correlated ferromagnetic phases in van der Waals heterostructures,\cite{Kurebayashi2022} and mark a first step towards a solid-state Bell test.\cite{Klobus2014} 

A false color scanning electron microscopy (SEM) image of the presented CPS device is shown in figure~\ref{fig:Fig_1}d. An indium arsenide (InAs) nanowire (NW) is contacted in the center by a titanium/aluminium superconducting source contact (S) and at the ends by titanium/gold normal metal contacts (N$_1$ and N$_2$), with a QD forming in each of the NW segments in between. Spin filtering is implemented by individually Zeeman splitting the QD resonances by the magnetic stray field that develops in the narrow gap fabricated in a long strip of a ferromagnetic metal (Permalloy), which we call ferromagnetic split-gate (FSG).\cite{Fabian2016,Bordoloi2020} We use the labels $(+,+)$ and $(-,-)$ for the two parallel and $(+,-)$ and $(-,+)$ for the two anti-parallel magnetization states. In the extended data figures 1 and 2, we demonstrate that the stray fields do not suppress superconductivity in the nearby S contact.

For later use, we introduce a spin polarization for each QD:\cite{Bordoloi2020} the Zeeman splitting of the QD states results in a spin-dependent single particle transmission (or transmission density of states) $D_{\sigma}$ for the spin state $\sigma  \in \{\uparrow,\downarrow\}$,\cite{Fabian2016,Desjardins2019,Bordoloi2020,Jiang2020} as illustrated in figure~\ref{fig:Fig_1}c. We define the spin polarization at the Fermi energy, $E_{\textrm{F}}$, for a given QD as
\begin{equation}
	\label{equ:nonideal_meas_op}
	P = \left. \frac{D_{\uparrow} - D_{\downarrow}}{D_{\uparrow}  + D_{\downarrow}} \right|_{E = E_{\rm F}}	
\end{equation} which depends on the stray magnetic field $B_{\textrm{str}}$, the external magnetic field $B$ and the gate voltage $V_{\textrm{g}}$. This polarization directly results in a spin-polarized current through the individual QDs.

First, we establish Cooper pair splitting by demonstrating a positive {\it charge} current cross-correlation, which corresponds to the simultaneous emission of two electrons from a Cooper pair into the different QDs. As illustrated in figure~\ref{fig:Fig_1}d, we simultaneously measure the two differential conductances $G_{j}=dI_j/dV$, with $I_j$ the current measured through QD $j$ and $V$ the common bias voltage applied to the superconductor S. The two conductances at zero dc bias and zero external magnetic field are plotted as a function of the FSG gate voltages $V_{\textrm{g1}}$ and $V_{\textrm{g2}}$ in figures~\ref{fig:Fig_1}e and \ref{fig:Fig_1}f, respectively. Each QD exhibits Coulomb blockade resonances tuned almost exclusively by the corresponding FSG voltage, suggesting very small cross capacitances to the respective far QD. In the following experiments, we avoid gate regions with parametric charge rearrangements, seen as small shifts in the resonance positions in the large area scans of figure~\ref{fig:Fig_1}f. Transport through one QD is only possible either mediated by Andreev reflection, or by higher order processes, since a double occupancy of the QD is suppressed by Coulomb interactions, and sequential tunneling through the QD by the required generation of a quasi-paricle in the superconductor.\cite{Fueloep2014} In our case the QDs are rather strongly coupled to S, so that the former process seems more probable (see extended data figure 2).

\begin{figure}[H]
	\centering
	\includegraphics[width=0.9\columnwidth]{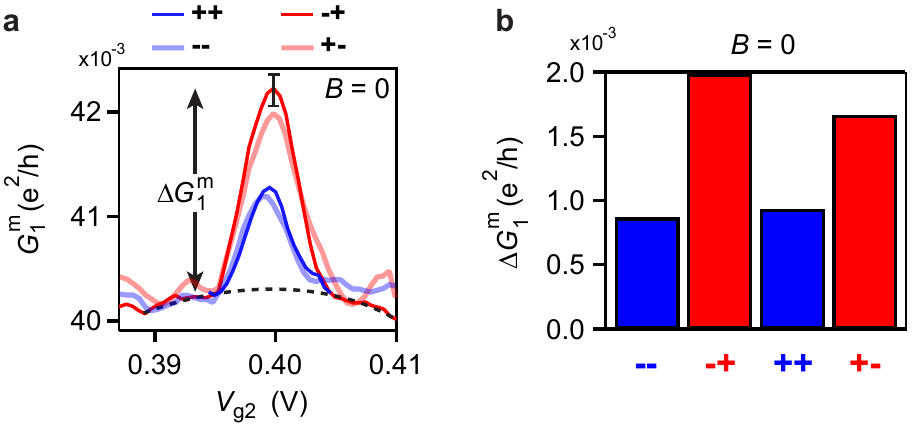}
	\caption{\textbf{Spin Correlation Measurements.} \textbf{a} Maximum conductance $G_{1}^{\textrm{m}}$ as a function of the gate voltage $V_{\textrm{g2}}$ for all four magnetization states at $B=0$ and $V_{\textrm{dc}} = 0$ for QD2 resonance R$_{2}$, showing a suppression of the conductance for the parallel magnetization states relative to the antiparallel states. The black dotted line shows the background for the state (-,+) subtracted in \textbf{b}. \textbf{b} Modulation amplitude of the maximum conductance $\Delta G^{\textrm{m}}_{\textrm{1}}$ for all four magnetization states extracted from \textbf{a}.}
	\label{fig:Fig_2}
\end{figure} 

If both QDs are resonant, i.e. at gate voltages at which a resonance of QD1 `crosses' one of QD2, Cooper pair splitting becomes allowed, and an additional current flows through {\it both} QDs. This is weakly visible in the large scale data of figure~\ref{fig:Fig_1}e. To see it better, we plot in figure~\ref{fig:Fig_1}g the resonance maximum $G^{\textrm{m}}_{1}$ of QD1 resonance B$_1$ as a function of $V_{{\rm g}2}$ (red), which tunes QD2 through the resonances A$_2$ to D$_2$, as plotted in black. $G^{\textrm{m}}_{1}$ shows pronounced peaks whenever QD2 becomes resonant, which can be directly attributed to the CPS process.\cite{Hofstetter2009,Schindele2012,Fueloep2014} To quantify this effect, we measure the modulation amplitude $\Delta G^{\textrm{m}}_{1}$, as indicated in figure~\ref{fig:Fig_1}g. Similar data for the other resonance A$_1$ is shown in extended data figure 3. This amplitude modulation vanishes if the superconductivity is suppressed by an external magnetic field, as demonstrated in extended data figure 4.

The main step now is to assess the spin correlations. To do this, we measure the charge correlations for the four different spin filter settings, i.e. for the four orientations of the FSGs. The sweep sequence to obtain these magnetization states is described in the Methods. Figure~\ref{fig:Fig_2}a shows the maximum conductance $G^{\textrm{m}}_{1}$ as a function of $V_{\textrm{g2}}$ for the two parallel and two anti-parallel FSG magnetization orientations at zero external magnetic field, $B=0$, for the resonances R$_{1}$ and R$_{2}$ shown in extended data figure 5. For all magnetizations we find a maximum in $G^{\textrm{m}}_{1}$ at the gate voltage of QD2 resonance R$_{2}$, demonstrating that the resonance positions do not change significantly with changes in the stray fields. The main finding is that the maximum conductance for the two parallel orientations is significantly reduced compared to the two anti-parallel orientations. This finding becomes even clearer in figure~\ref{fig:Fig_2}b, where we plot the modulation of the maximum conductance, $\Delta G^{\rm m}_1$, for the four magnetization states. To obtain this value, we individually subtract a parabolic background for each curve, shown as black dashed line for the state $(-,+)$ in figure~\ref{fig:Fig_2}a. Figure~\ref{fig:Fig_2}b directly illustrates that the modulation amplitude is suppressed by a factor of $\sim 2$ for the two parallel magnetization states with respect to the two anti-parallel states, in qualitative agreement with the negative spin cross-correlation expected for Cooper pair emission.

To discriminate the CPS contribution from other transport processes, we now investigate the conductance amplitude modulation in both arms, for which we first show that $\Delta G^{\textrm{m}}_{1}\approx \Delta G^{\textrm{m}}_{2}$, as expected for CPS, and that in both arms the conductance maxima are larger for the anti-parallel

\begin{figure}[H]
	\centering
	\includegraphics[width=0.95\columnwidth]{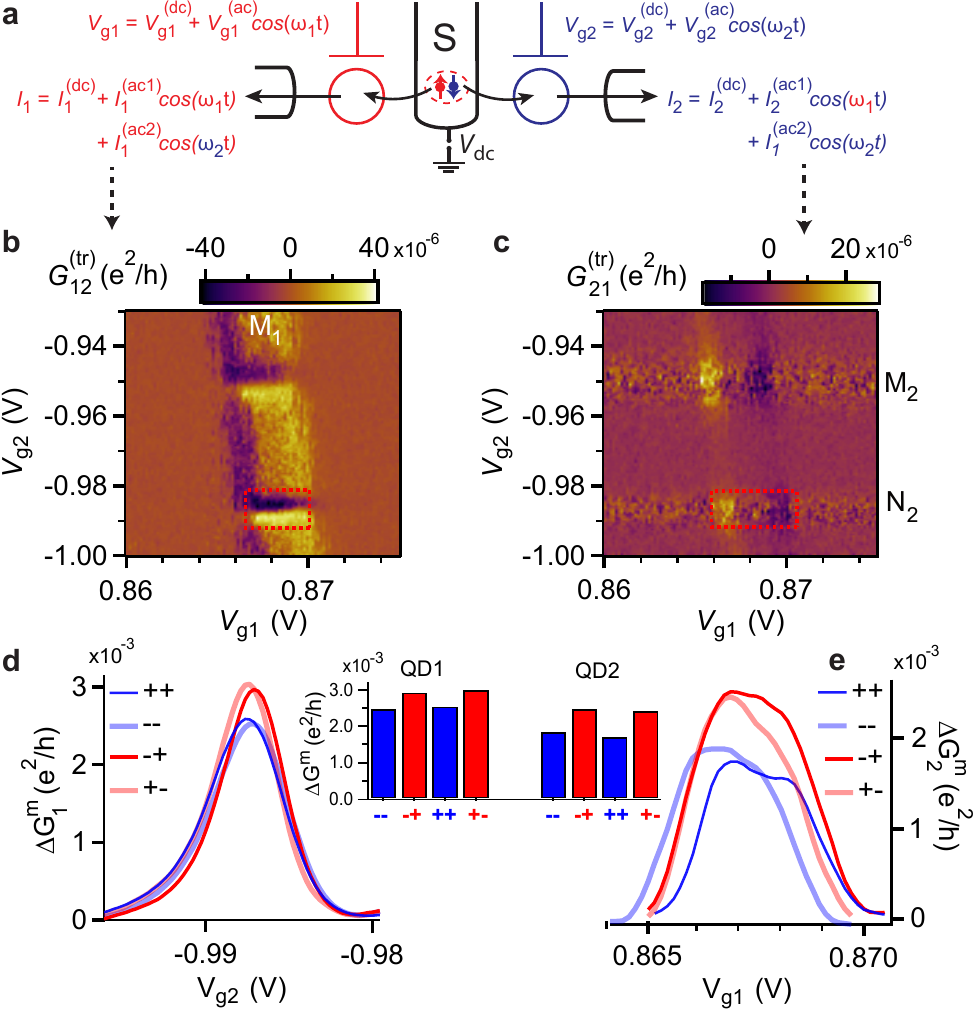}
	\caption{\textbf{Transconductance measurements.} \textbf{a} Illustration of the transconductance measurements in a CPS device. \textbf{b,c} Transconductance $G^{\textrm{(tr)}}_{12} = \frac{I_{1}^{\textrm{(ac2)}}}{V_{\textrm{g2}}^{\textrm{(ac)}}}$ and $G^{\textrm{(tr)}}_{21} = \frac{I_{2}^{\textrm{(ac1)}}}{V_{\textrm{g1}}^{\textrm{(ac)}}}$, respectively, plotted as a function of $V_{\textrm{g1}}$ and $V_{\textrm{g2}}$ for resonance $\textrm{M}_{1}$ in QD1 and resonances $\textrm{M}_{2}$ and $\textrm{N}_{2}$ in QD2 at $B=0$ for the (+,-) magnetization state. \textbf{d,e} Modulation of the maximum conductance $\Delta G_{1}^{\textrm{m}}$ and $\Delta G_{2}^{\textrm{m}}$, respectively, reconstructed from \textbf{b} and \textbf{c} (see Methods) for all four magnetization states at $B=0$. \textbf{Inset:} Bar plot of the maximum conductance modulation $\Delta G^{\textrm{m}}$ for QD1 and QD2 for the four magnetization states.}
	\label{fig:Fig_3}
\end{figure}

\noindent magnetization configurations than for the parallel, only expected for the CPS mechanism (the spin-dependence of other transport mechanisms is discussed in extended data table 1). In direct experiments as presented above, we can resolve the CPS current only in QD1, but not in QD2, due to the large background conductance. To resolve the CPS currents in both arms, we perform transconductance experiments, as illustrated in figure~\ref{fig:Fig_3}a and described in details in the Methods section. The basic concept is to measure the modulation of the current in one QD upon modulating the gate voltage of the other QD, which is only sensitive to processes in which both QDs are involved.

We thus measure the transconductances $G^{\textrm{(tr)}}_{12} \coloneqq \frac{dI_{1}}{dV_{\textrm{g2}}}=\frac{I_{1}^{\textrm{(ac2)}}}{V_{\textrm{g2}}^{\textrm{(ac)}}}$ and $G^{\textrm{(tr)}}_{21} \coloneqq \frac{dI_{2}}{dV_{\textrm{g1}}}=\frac{I_{2}^{\textrm{(ac1)}}}{V_{\textrm{g1}}^{\textrm{(ac)}}}$ (with the second step relating to the lock-in detection scheme described in figure~\ref{fig:Fig_3}a and the Methods). Both transconductances are plotted as a function of the gate voltages $V_{\rm g1}$ and $V_{\rm g2}$ in figures~\ref{fig:Fig_3}b and ~\ref{fig:Fig_3}c, respectively. Both exhibit peak-valley features at the indicated QD resonances, consistent with maxima in $G_1$ and $G_2$ when measured along the respective resonances. The direct integration of these data allows us to reconstruct the modulation of the conductance maxima $\Delta G_{1}^{\rm m}$ and $\Delta G_{2}^{\rm m}$ (details in Methods). The results are plotted in figures~\ref{fig:Fig_3}d and ~\ref{fig:Fig_3}e for all four magnetization states at $B=0$, for the resonance crossings $\textrm{M}_{1}$ and $\textrm{N}_{2}$ in the gate voltage intervals indicated by the dashed rectangle in figures~\ref{fig:Fig_3}b and ~\ref{fig:Fig_3}c, respectively (see extended data figure 6 for details).

The reconstructed conductance modulations are similar as in the previously discussed data set and of similar magnitude for both QDs, in spite of rather different background conductances. With both conductance variations available, we can assess the fraction of CPS in the signals. We expect a positive variation in both arms only for CPS, so that we take the minimum of both signals as a lower bound for CPS,\cite{Schindele2012} i.e. $\Delta G^{\textrm{cps}} = \Delta G_{2}^{\textrm{m}}$ as $\Delta G_{1}^{\textrm{m}} > \Delta G_{2}^{\textrm{m}}$, while the large backgrounds in both signals that depend only on one QD are due to local pair tunneling (LPT), and the remaining part $\Delta G^{\textrm{other}} = \Delta G_{1}^{\textrm{m}} - \Delta G_{2}^{\textrm{m}}$ is due to CPS followed by direct cotunneling between the QDs, as discussed in extended data table 1. The fraction of the CPS generated currents in our devices are on a similarly small scale as in previous NW-based experiments, which we express as the efficiency\cite{Schindele2012} $\eta_{\rm tot} = \frac{2\Delta G^{\rm cps}}{G_1^{\rm m}+G_2^{\rm m}}\approx 3\%$. However, the CPS fraction in the signals that depend on \textit{both} QDs is very large, namely $\eta_{\rm 2dot}=\frac{\Delta G^{\textrm{cps}}}{\Delta G^{\textrm{cps}}+\Delta G^{\textrm{other}}} \approx 85\%$. This latter number suggests that most of the transconductance is due to CPS.

Most importantly, we again show the maximum modulations for the four magnetization states as bar plots in the shared inset of figures~\ref{fig:Fig_3}d and ~\ref{fig:Fig_3}e, demonstrating the {\it simultaneous} suppression of the signals in both arms for the parallel magnetization states compared to the anti-parallel states, directly illustrating a spin anti-correlation. We show explicitly in the Methods how we extract the expectation value of the spin correlation operator from the experiments. The crucial steps are that we first account for non-ideal QD polarizations, $P_{j} < 1$, meaning for example that an electron in state $\ket{\uparrow}$ can enter the detector set to the opposite magnetization and would be wrongly detected as a $\ket{\downarrow}$ electron. To do this, we introduce the normalized tunnel probabilities at the Fermi energy $\Gamma_{\uparrow,\downarrow}^{i} = \left. \frac{D_{\uparrow,\downarrow}}{D_{\uparrow} + D_{\downarrow}} \right|_{E = E_{\rm F}}	$ for each spin species and FSG magnetization $i \in \{+,-\}$, such that the non-ideal projection operators is given by $\hat{M}^{i}_{\textrm{nid}} = \Gamma_{\uparrow}^{i} \ket{\uparrow}\bra{\uparrow} + \Gamma_{\downarrow}^{i} \ket{\downarrow}\bra{\downarrow}$ with $\Gamma_{\uparrow}^{i} + \Gamma_{\downarrow}^{i} = 1$. With these non-ideal projection operators, we define the non-ideal (nid) spin cross-correlation operator\cite{Klobus2014}
\begin{equation}
	\label{equ:nonideal_meas_op}
	\hat{C}_{\textrm{nid}} = (\hat{M}^{+}_{1,\textrm{nid}} - \hat{M}^{-}_{1,\textrm{nid}}) \otimes (\hat{M}^{+}_{2,\textrm{nid}} - \hat{M}^{-}_{2,\textrm{nid}}).	
\end{equation}
For example, the expectation value for ideally split Cooper pairs reads $\bra{\textrm{cps}}\hat{C}_{\textrm{nid}}\ket{\textrm{cps}} = - P_{1}P_{2} < 0$, with $P_1$ and $P_2$ the magnitudes of the spin filter polarizations. 

Next, we use the tunnel probabilities to express the conductance variations and $\hat{C}_{\rm nid}$ (see Methods for details), with the intuitive result that we can use the conductance variations in the four magnetization states to estimate the spin correlation in the experiments:
\begin{equation}
	\label{equ:corr_value}
	\left<\hat{C}\right>_{\textrm{exp}} \coloneqq \frac{\Delta G_{++}^{\textrm{cps}} - \Delta G_{+-}^{\textrm{cps}}}{\Delta G_{++}^{\textrm{cps}} + \Delta G_{+-}^{\textrm{cps}}},
\end{equation}
where $\Delta G_{++}^{\textrm{cps}}$ and $\Delta G_{+-}^{\textrm{cps}}$ are the measured CPS conductance variations. We note that based on the results above, we neglect here non-CPS contributions involving both QDs. For the data set of figure~\ref{fig:Fig_2}b, we find a spin cross-correlation of $\left<\hat{C}\right>_{\textrm{exp}}\approx -0.37$, clearly demonstrating a negative correlation between the spin signals. Assuming that only CPS contributes to the conductance variations, we obtain the geometric mean of the polarizations as $\bar{P} = \sqrt{P_1P_2}\approx 60\%$ at $B = 0$ and on resonance, similar to values achieved in a double quantum dot spin valve.\cite{Bordoloi2020}
Similarly, we find for the data in figure~\ref{fig:Fig_3}d and \ref{fig:Fig_3}e the correlation $\left<\hat{C}\right>_{\textrm{exp}} \sim -0.12$ and $\bar{P}\approx 35\%$, smaller than for the resonances in figure~\ref{fig:Fig_2}, consistent with a reduced mean polarization due to a larger life time broadening of these QD states.\cite{Bordoloi2020}

\begin{figure}[H]
	\centering
	\includegraphics[width=\columnwidth]{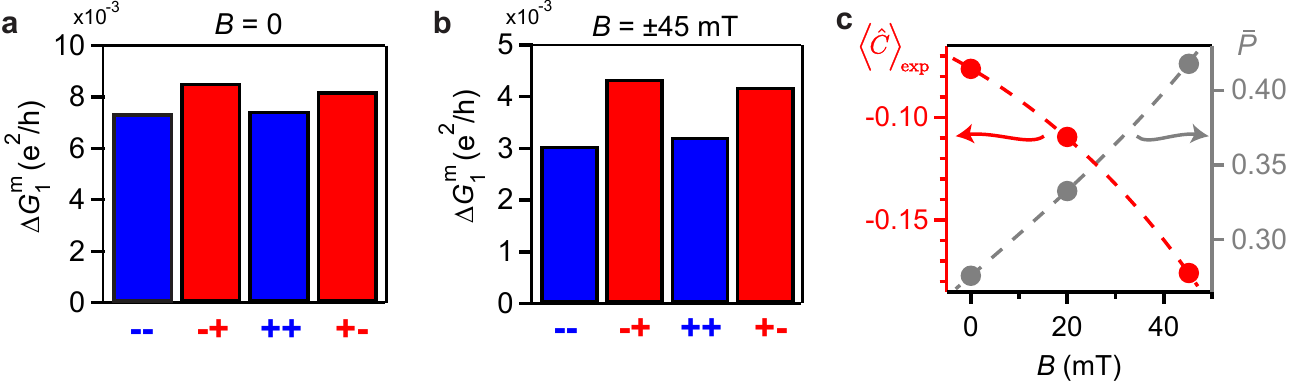}
	\caption{\textbf{Magnetic field tuning of the QD spin polarization.} \textbf{a,b} Modulation of the conductance maximum $\Delta G^{\textrm{m}}_{\textrm{1}}$ for all four magnetization states at $B=0$ and $B = \pm 45\,$mT, respectively, with the (-,-) and (+,-) configurations measured at $B = -45\,$mT, and the (+,+) and (-,+) configurations at $B = +45\,$mT. \textbf{c} Experimental spin cross-correlation $\langle \hat{C} \rangle _{\textrm{exp}}$ (red) and the extracted geometric mean $\bar{P}$ of the QD spin polarizations (grey) as a function of the external magnetic field $B$, showing a stronger negative correlation for larger QD polarizations.} 
	\label{fig:Fig_4}
\end{figure}

The mean QD polarization and the spin filtering effect of a QD, which consequently influences $\left<\hat{C}\right>_{\textrm{exp}}$, can be further increased by applying a homogeneous external magnetic field, limited by the larger switching field to still be able to access all magnetization states,\cite{Bordoloi2020} and by the critical magnetic field of the superconductor. Figures~\ref{fig:Fig_4}a,b show the measured modulation in the conductance maximum $\Delta G^{\textrm{m}}_{1}$ for the resonances X$_{1}$ and X$_{2}$ shown in extended data figure 7, with their

\noindent respective backgrounds subtracted, and for all four magnetization states at $B=0$ and $B = \pm45\,$mT, respectively (CPS signals for $B = \pm20\,$mT can be found in extended data figure 7). $\Delta G^{\textrm{m}}_{1}$ for the (-,-) and (+,-) states were measured at $B=-45\,$mT, while the ones for (+, +) and (-,+) were measured at  $B=+45\,$mT, so that the external field boosts the stray field magnitude of the wider, already switched FSG. We again find that the maximum modulations are suppressed for both parallel states with respect to the two anti-parallel states, consistent with a negative spin cross-correlation between the split Cooper pair electrons. We note that all absolute modulation amplitudes $\Delta G^{\textrm{m}}_{1}$ decrease with increasing $B$, as does the single QD backgrounds (see extended data figure 8), qualitatively consistent with local pair tunneling and spin filtering in one QD. For a given $B$, we extract the expectation value of the spin correlation operator $\langle \hat{C} \rangle _{\textrm{exp}}$ using equation~\ref{equ:corr_value}, which is plotted in figure~\ref{fig:Fig_4}c (red symbols). Again assuming an ideal CPS process, we find that the spin cross-correlation becomes stronger and more negative for increasing $B$, increasing from $\langle \hat{C} \rangle _{\textrm{exp}} \approx -0.07$ at $B=0$ to $\langle \hat{C} \rangle _{\textrm{exp}} \approx -0.17$ at $B=45\,$mT, consistent with a larger mean QD spin polarization induced by the additional Zeeman splitting in the external magnetic field. The corresponding mean polarization $\bar{P}$ extracted from $\langle \hat{C} \rangle _{\textrm{exp}} = -P_{1}P_{2} = -\bar{P}^{2}$, is plotted in figure~\ref{fig:Fig_4}c (grey symbols) on the right axis. As one might expect, the mean polarization increases linearly, from $\bar{P} \approx 27\%$ at $B=0$ to $\bar{P} \approx 42\%$ at $B=45\,$mT, consistent with Zeeman split QD states. These polarization values could be improved either by stronger stray fields, for example using other FSG materials like Cobalt, or by increasing the QD lifetime,\cite{Thomas2020} such that spin cross-correlation experiments with close to ideal anti-correlations $\langle \hat{C}\rangle \approx -1$ become feasible.

As a last step we critically assess one major effect that might influence the spin correlation in our experiments: the FSG stray fields at the position of the superconductor could result in a suppression of the CPS signals. However, even neglecting the Meissner effect preventing magnetic fields from penetrating S, we can expect considerably larger stray fields on S in the anti-parallel states, compared to the parallel ones,\cite{Maurer2018} which would suppress superconductivity effects like CPS stronger for the anti-parallel configuration, the exact opposite of our findings. As a control, we find that CPS and local pair tunneling are both reduced in a homogeneous external magnetic field (extended data figure 8).

In conclusion, we demonstrate spin correlation experiments in an electron entangler device based on Cooper pair splitting from a superconductor, for which we find a negative value for the spin cross-correlation, $\langle \hat{C} \rangle_{\textrm{exp}} \approx -0.4<0$, as expected for split spin singlet Cooper pairs. In novel transconductance measurements, we find similar magnitudes for the CPS conductances in both arms of the entangler, and a similar suppression of the CPS signals for parallel orientations of the spin filters compared to the anti-parallel, in accordance with Cooper pair splitting. In addition, we report CPS measurements in finite external magnetic fields and demonstrate that the cross correlation signals deviate from the ideal value mainly due to finite QD spin polarizations, not due to spurious transport effects. These findings suggest that in devices with stronger QD confinement, it should be possible to detect the full spin anti-correlation. Our work, general enough to be implemented in any solid-state system with reasonably discrete states, can be used to investigate many fundamental states of matter related to the electron spin, such as topological, unconventional or triplet superconductivity,\cite{Bergeret2005,Benjamin2006,Jeon2021} equal spin Andreev reflection\cite{He2014} in hybrid Majorana\cite{Mourik2012,Deng2016} devices, or various correlated ferromagnetic phases, for example in van der Waals heterostructures.\cite{Kurebayashi2022} Since these entanglers rely on single electron tunneling, they can in principle be operated on the single Cooper pair level,\cite{Hollosy2015,Ranni2022} ideally suited to perform Bell tests with electron spins.\cite{Braunecker2013,Klobus2014}

\section*{METHODS}

\subsection{Conductance and Transconductance Measurements}\hfill

\noindent The differential conductance for each QD, $G_{i} = \textrm{d}I_{i}/\textrm{d}V$ with $i=1,2$ and $V = V_{\textrm{dc}} + V_{\textrm{ac}} \cos(\omega t)$,  were measured simultaneously using standard lock-in techniques with a modulation of the bias voltage of $V_{\rm ac} = 10\,\mu$V, at a frequency of 177.77 Hz and at a temperature of $\sim 50\,$mK.

\noindent For the transconductance measurements, we apply a dc voltage of $V_{\rm dc} = 25\,\mu$V to S and measure the variation in the current in one arm that originates from the modulation of the gate voltage of the other arm, which ideally corresponds to the derivative along a QD resonance.

\noindent In practice, we modulate the FSG2 voltage on QD2 by an ac voltage $V_{\textrm{g2}}^{\textrm{(ac)}} = 0.5\,$mV at the frequency $\omega_{2} = 37.77\,$Hz and measure the resulting modulation in the QD1 current, $I_1^{\rm (ac2)}$, at the same frequency $\omega_{2}$. The resulting  transconductance $G^{\textrm{(tr)}}_{12} = \frac{I_{1}^{\textrm{(ac2)}}}{V_{\textrm{g2}}^{\textrm{(ac)}}}$ is plotted in figure~\ref{fig:Fig_3}b as a function of the gate voltages $V_{\rm g1}$ and $V_{\rm g2}$. Similarly, we plot in figure~\ref{fig:Fig_3}c the transconductance $G^{\textrm{(tr)}}_{21} = \frac{I_{2}^{\textrm{(ac1)}}}{V_{\textrm{g1}}^{\textrm{(ac)}}}$ that describes the response of the QD2 current to a modulation of $V_{\textrm{g1}}^{\textrm{(ac)}} = 0.5\,$mV in $V_{\rm g1}$ at a frequency $\omega_1 = 77.77\,\textrm{Hz} \neq \omega_2$.

\noindent To reconstruct the the amplitude modulation for each resonance, we first average five cross sections centered on a resonance maximum, as indicated by the red dashed rectangles in figures~\ref{fig:Fig_3}b and ~\ref{fig:Fig_3}c. From this curve (not shown), we reconstruct $\Delta G_{1,2}^{\textrm{m}}$ at the resonance crossing by integrating the transconductance over the voltage on the far gate: 
\begin{equation}
	\label{equ:trans_cond}
	\Delta G_{j}^{\textrm{m}} = \frac{1}{V_{\textrm{dc}}} \int_{-\infty}^{V_{\textrm{g}k}} G_{jk}^{\rm (tr)} (\tilde{V}_{\textrm{g}k}) d\tilde{V}_{\textrm{g}k}, 
\end{equation}
where $j$ and $k \neq j$ refer to arms 1 and 2, respectively. The obtained modulation of the conductance maxima $\Delta G_{1}^{\textrm{m}}$ and $\Delta G_{2}^{\textrm{m}}$ are plotted in figures~\ref{fig:Fig_3}d and ~\ref{fig:Fig_3}e.

\subsection{Device Fabrication}\hfill

 \noindent The InAs NWs were grown by gold (Au) colloid assisted chemical beam epitaxy\cite{Gomes2015} and have a diameter of 50 $\pm$ 5 nm, depending on the size of the gold seed particle. The NWs were mechanically transferred from the growth substrate to a heavily p-doped silicon substrate serving as a global backgate (BG), with a 400 nm $\rm SiO_{2}$ insulating top layer. For the electron beam lithography, we employed pre-defined markers and contact pads made of Ti/Au (5 nm/ 45 nm). The normal metal contacts at the NW ends are made of Ti/Au (5 nm/ 45 nm) and the superconducting contact of Ti/Al (5 nm/ 70 nm). Two $5.5\,$nm wide InP barriers\cite{Thomas2020} separated by $19\,$nm InAs below S help to reduce the direct tunnel coupling between the two QD segments. The ferromagnetic split-gates are $6\,\mu$m long on both sides of the NW and are 100 nm and 180 nm wide, respectively, and made of 30 nm thick Permalloy (Py).\cite{Fabian2016,Bordoloi2020} Before evaporating the normal metal and superconducting contact materials, the native oxide of the NW was etched with a 1:10 ratio $(\rm NH_{4})_{2}S_{x} : H_{2}O$ solution for 3.5 minutes. The $(\rm NH_{4})_{2}S_{x}$ solution was prepared by mixing 0.96 grams of sulfur powder in 10 ml of ammonium sulfide solution (20$\%$ in $\rm H_{2}O$).

\subsection{Magnetic Measurements}\hfill

\noindent Each FSG generates a strongly localized stray field, $B_{\rm str}$, at the corresponding QD position, in the direction of the strip magnetization, which is oriented either parallel or anti-parallel to the long strip axis due to the shape anisotropy.\cite{Fabian2016,Bordoloi2020,Aurich2010} The FSG magnetization, and with it $B_{\rm str}$, can be inverted at a characteristic external  switching field, $B_{\rm sw}$, determined by the FSG width in the device design, demonstrated previously using anisotropic magnetoresistance (AMR) measurements\cite{Aurich2010} as well as spin valve measurements.\cite{Bordoloi2020} From independent experiments discussed in extended data figure 9, we obtain $B_{{\rm sw}1}\approx100\,$mT for FSG1 and $B_{{\rm sw}2}\approx25\,$mT for FSG2, which allows us to access all four relative FSG magnetization orientations by specific sweep sequences of the external magnetic field $B$ applied along the FSG axes, as discussed below. We label these magnetization states $(+,+)$ and $(-,-)$ for the two parallel and $(+,-)$ and $(-,+)$ for the two anti-parallel configurations.

\subsection{Spin cross-correlation operator for non-ideal spin detectors}\hfill

\noindent Here, we derive the spin cross-correlation operator for non-ideal spin detectors and the relation  to the differential conductances measured in a Cooper pair splitter for the four magnetization states of the ferromagnetic sidegates (FSGs). 

\noindent In a Cooper pair splitter with {\it ideal} transmission and {\it ideal} spin projection operators $M_{k}^{+}=\ket{\uparrow}\bra{\uparrow}$ and $M_{k}^{-}=\ket{\downarrow}\bra{\downarrow}$ in each arm $k \in \{ 1,2 \} $, the current in one arm is $\propto ( M_{k}^{+}+M_{k}^{-})$ and the spin current $\propto ( M_{k}^{+}-M_{k}^{-} )$. The spin current correlation operator between the two arms thus reads
\begin{equation}
	\label{equ_spin_corr_ideal}
	\begin{split}
		\hat{C}_{\textrm{ideal}} & = \left( M_{1}^{+}-M_{1}^{-} \right) \otimes \left( M_{2}^{+}-M_{2}^{-} \right)\\
		&= \ket{\uparrow \uparrow}\bra{\uparrow \uparrow} - \ket{\uparrow \downarrow}\bra{\uparrow \downarrow} - \ket{\downarrow \uparrow}\bra{\downarrow \uparrow} + \ket{\downarrow \downarrow}\bra{\downarrow \downarrow}.
	\end{split}
\end{equation}

\noindent For the expectation value of $\hat{C}_{\textrm{ideal}}$ for Cooper pairs  $\ket{\textrm{cps}} = \frac{1}{\sqrt{2}} [\ket{\uparrow \downarrow} - \ket{\downarrow \uparrow}]$ emitted in the CPS device one directly finds
\begin{equation}
	\label{equ:spin_corr_ideal_exp}
	\bra{\textrm{cps}}\hat{C}_{\textrm{ideal}}\ket{\textrm{cps}} = -1,
\end{equation}
while for uncorrelated electrons $\left<\hat{C}_{\textrm{ideal}}\right> = 0$.

\noindent To account for {\it non-ideal} spin detectors and transmissions, we define the spin polarization of a spin filter - in our case a QD - at the Fermi energy $E_{\rm F}$ as
\begin{equation}
	\label{equ:DefPol}
		P = \left. \frac{D_{\uparrow} - D_{\downarrow}}{D_{\uparrow}  + D_{\downarrow}} \right|_{E = E_{\rm F}}
\end{equation} 
with $D_{\sigma}$ the QD transmission (or transmission density of states) for spin state $\sigma  \in \{\uparrow,\downarrow\}$.
For non-ideal spin detectors with $P_{j}<1$, an electron in state $\ket{\uparrow}$ can enter the detector set to detect $\ket{\downarrow}$, and thus get wrongly detected as $\ket{\downarrow}$ with non-zero probability. To accommodate such processes, we define the non-ideal measurement operators of a spin detector as\cite{Klobus2014}
\begin{equation}
	\label{equ:nonideal_meas_op}
	\begin{split}
		\hat{M}^{+}_{\textrm{nid}} = \Gamma_{\uparrow}^{+} \ket{\uparrow}\bra{\uparrow} + \Gamma_{\downarrow}^{+} \ket{\downarrow}\bra{\downarrow}\\
		\hat{M}^{-}_\textrm{nid}{} = \Gamma_{\uparrow}^{-} \ket{\uparrow}\bra{\uparrow} + \Gamma_{\downarrow}^{-} \ket{\downarrow}\bra{\downarrow},
	\end{split}
\end{equation}
where +(-) denotes the detector set parallel (antiparallel) to the given quantization axis and $\Gamma_{\uparrow,\downarrow}^{j}$ is given by
\begin{equation}
	\label{equ:cps_gamma}
	\Gamma_{\uparrow,\downarrow}^i = \left. \frac{D_{\uparrow,\downarrow}^i}{D_{\uparrow}^i + D_{\downarrow}^i} \right|_{E = E_{\rm F}} = \left. \frac{D_{\uparrow,\downarrow}^i}{D_{\rm tot}^i} \right|_{E = E_{\rm F}}
\end{equation} 
with $i \in \{+,-\}$ and $\Gamma_{\uparrow}^{i} + \Gamma_{\downarrow}^{i} = 1$. Using the spin polarization from Eq.~\ref{equ:DefPol}, we can express $\Gamma_{\uparrow,\downarrow}^{i}$ as
\begin{equation}
	\label{equ:cps_gamma_pol}
	\begin{split}
		\Gamma_{\uparrow}^{i} = \frac{1}{2} (1 + P^{i}) \,\,\text{and}\,\, \Gamma_{\downarrow}^{i} = \frac{1}{2} (1 - P^{i})
	\end{split}
\end{equation}
with $P^i$ the polarization for this QD in magnetization state $i$.
For example, for fully polarized spin detectors, $P^{+}=1$ and $P^{-}=-1$, so that $\Gamma_{\uparrow}^{+} = \Gamma_{\downarrow}^{-} = 1$ and $\Gamma_{\uparrow}^{-} = \Gamma_{\downarrow}^{+} = 0$. From Eq.~\ref{equ:nonideal_meas_op}, one then again obtains the ideal spin projection operators $\hat{M}^{+}_{\textrm{ideal}} = \ket{\uparrow}\bra{\uparrow}$ and $\hat{M}^{-}_{\textrm{ideal}} = \ket{\downarrow}\bra{\downarrow}$. 
Similarly using Eq.~\ref{equ:nonideal_meas_op}, one obtains the non-ideal spin current operator for a single QD, 
\begin{equation}
	\label{equ:cps_spinproj_single}
	\begin{split}
		\hat{M}^{+}_{\textrm{nid}} - \hat{M}^{-}_{\textrm{nid}} & = (\Gamma_{\uparrow}^{+} - \Gamma_{\uparrow}^{-}) \ket{\uparrow}\bra{\uparrow} + (\Gamma_{\downarrow}^{+} - \Gamma_{\downarrow}^{-})\ket{\downarrow}\bra{\downarrow}\\
		& =  P  (\ket{\uparrow}\bra{\uparrow} - \ket{\downarrow}\bra{\downarrow})\\
		& =  P  (\hat{M}^{+}_{\textrm{ideal}} - \hat{M}^{-}_{\textrm{ideal}}),
	\end{split}
\end{equation} 
where we assume for the second step that the magnitude of the spin polarizations in the spin filter settings are equal, $P^{+} = -P^{-} = P$.
Now we can write down the spin current cross correlation operator,
\begin{equation}
	\label{equ:cps_spinproj_nonideal}
	\begin{split}
		\hat{C}_{\textrm{nid}} &= (\hat{M}^{+}_{1,\textrm{nid}} - \hat{M}^{-}_{1,\textrm{nid}}) \otimes (\hat{M}^{+}_{2,\textrm{nid}} - \hat{M}^{-}_{2,\textrm{nid}})\\
		& = P_{1} (\ket{\uparrow}\bra{\uparrow} - \ket{\downarrow}\bra{\downarrow}) \otimes P_{2} (\ket{\uparrow}\bra{\uparrow} - \ket{\downarrow}\bra{\downarrow})\\ 
		&= P_{1} P_{2} (\ket{\uparrow \uparrow}\bra{\uparrow \uparrow} - \ket{\uparrow \downarrow}\bra{\uparrow \downarrow} - \ket{\downarrow \uparrow}\bra{\downarrow \uparrow} + \ket{\downarrow \downarrow}\bra{\downarrow \downarrow})\\
		& = P_{1} P_{2} \hat{C}_{\textrm{ideal}}
	\end{split}
\end{equation}
where $P_{k}^{+} = -P_{k}^{-} = P_{k}$ are the polarization magnitudes of the two spin detectors,  QD1 and QD2 in our case.

\noindent For illustration, we quickly discuss some simple cases: for a spin singlet CPS state, the expectation value of the non-ideal spin cross-correlation operator is $\bra{\rm cps}\hat{C}_{\textrm{nid}}\ket{\rm cps} = - P_{1} P_{2} < 0$, illustrating the opposite spins of the split Cooper pair electrons. For fully polarized spin filters, $P_{1} = P_{2} = 1$, one obtains $\bra{\rm cps}\hat{C}_{\textrm{nid}}\ket{\rm cps} = -1$, i.e. the experiment would show a full spin anti-correlation. This ideal value is reduced by non-ideal polarizations, in spite of the CPS state being fully anti-correlated. This shows that the detector quality is relevant for estimating the correlation value in such experiments. 

\noindent Another instructive example is the product state $\ket{s} = \ket{\uparrow \downarrow}$, which as well yields $\bra{\uparrow \downarrow}\hat{C}_{\textrm{nid}}\ket{\uparrow \downarrow} = - P_{1} P_{2} < 0$, the same as the $\ket{\rm cps}$ state. In contrast, the product state $\ket{s} = \ket{\uparrow \uparrow}$ results in $\bra{\uparrow \uparrow}\hat{C}_{\textrm{nid}}\ket{\uparrow \uparrow} = P_{1} P_{2} > 0$, demonstrating a positive correlation. To discriminate the CPS state  from the product state $ \ket{\uparrow \downarrow}$, one would need to perform additional spin cross-correlation measurements with the quantization axis along the other orthogonal axes \^{x},\^{y}, as suggested as a Bell test for electrons.\cite{Klobus2014,Braunecker2013} For all directions, we expect $\bra{\rm cps}\hat{C}_{\textrm{nid},\hat{\textrm{x}},\hat{\textrm{y}},\hat{\textrm{z}}}\ket{\rm cps} = - P_{1} P_{2}$ as before, because the $\ket{\rm cps}$ singlet state is isotropic. In contrast, the product state results in $\bra{\uparrow \downarrow}\hat{C}_{\textrm{nid},\hat{\textrm{x}}}\ket{\uparrow \downarrow} = \bra{\uparrow \downarrow}\hat{C}_{\textrm{nid},\hat{\textrm{y}}}\ket{\uparrow \downarrow} = 0$. Spin cross-correlation measurements along three orthogonal directions would therefore be sufficient to differentiate the maximally entangled $\ket{\textrm{cps}}$ state from other non-entangled states. Less stringent entanglement witness operators were proposed in References 16 and 32.

\noindent As a last step, we need to find a relation between the measured differential conductances and the expectation value of the correlator. The total conductance through a given arm with spin filter setting $i \in \{+,-\}$ is the sum over the two spin channels, which we separate using the spin projection operators:
\begin{equation}\label{equ:singleQD_con_operator}
	\begin{split}
		\hat{G}^{i}  & = \frac{e^{2}}{h} \left(D_{\uparrow}\ket{\uparrow}\bra{\uparrow} + D_{\downarrow} \ket{\downarrow}\bra{\downarrow}\right)\\
		& = \frac{e^{2}}{h} D_{\rm tot} \left(\Gamma_{\uparrow}^{i} \ket{\uparrow}\bra{\uparrow} + \Gamma_{\downarrow}^{j} \ket{\downarrow}\bra{\downarrow}\right).\\
	\end{split}	
\end{equation} 

\noindent The part of the conductance through one QD that depends on both dots, i.e. the {\it variation} of the QD resonance maxima in the main text, we describe for simplicity as the product of the individual QD transmissions, which in turn depend on the respective spin filter orientations $i,j \in \{+,-\}$:
\begin{equation}
	\label{equ:twoQD_con_operator}
	\begin{split}
		\Delta\hat{G}^{ij} & =\overbrace{\left(\frac{e^2}{h}\right)^2 D_{\rm tot,1}D_{\rm tot,2}}^{=:K}  \left[\Gamma_{1 \uparrow}^{i} \ket{\uparrow}\bra{\uparrow} + \Gamma_{1 \downarrow}^{i} \ket{\downarrow}\bra{\downarrow}\right] \otimes \left[\Gamma_{2 \uparrow}^{j} \ket{\uparrow}\bra{\uparrow} + \Gamma_{2 \downarrow}^{j} \ket{\downarrow}\bra{\downarrow}\right]\\
		& = K\left[\Gamma_{1 \uparrow}^{i} \Gamma_{2 \uparrow}^{j} \ket{\uparrow \uparrow}\bra{\uparrow \uparrow} + \Gamma_{1 \uparrow}^{i} \Gamma_{2 \downarrow}^{j} \ket{\uparrow \downarrow}\bra{\uparrow \downarrow} + \Gamma_{1 \downarrow}^{i} \Gamma_{2 \uparrow}^{j} \ket{\downarrow \uparrow}\bra{\downarrow \uparrow} + \Gamma_{1 \downarrow}^{i} \Gamma_{2 \downarrow}^{j} \ket{\downarrow \downarrow}\bra{\downarrow \downarrow}\right].
	\end{split}	
\end{equation}
We now combine these conductance variations into an expression containing the correlation operator, 
\begin{equation}
	\label{equ:conductance_operator}
	\begin{split}
		\Delta\hat{G}^{++} - \Delta \hat{G}^{+-} - \Delta\hat{G}^{-+} + \Delta\hat{G}^{--} & =K\cdot P_{1} P_{2} \left[\ket{\uparrow \uparrow}\bra{\uparrow \uparrow} - \ket{\uparrow \downarrow}\bra{\uparrow \downarrow} - \ket{\downarrow \uparrow}\bra{\downarrow \uparrow} + \ket{\downarrow \downarrow}\bra{\downarrow \downarrow}\right]\\
		&\equiv K \hat{C}_{\textrm{nid}}. 
	\end{split}	
\end{equation}
This combination of the conductance variations in the four magnetization states is therefore a measure for the spin cross-correlation. To account for the prefactor $K$, we also calculate the following combination:
\begin{equation}
	\label{equ:conductance_operator_add}
	\begin{split}
		\Delta\hat{G}^{++} + \Delta\hat{G}^{+-} + \Delta\hat{G}^{-+} + \Delta\hat{G}^{--}
		& = K  \underbrace{\left[\ket{\uparrow \uparrow}\bra{\uparrow \uparrow} + \ket{\uparrow \downarrow}\bra{\uparrow \downarrow} + \ket{\downarrow \uparrow}\bra{\downarrow \uparrow} + \ket{\downarrow \downarrow}\bra{\downarrow \downarrow}\right]}_{\textrm{= 1}}=K,
	\end{split}	
\end{equation}
independent of the QD polarizations. Dividing equation~\ref{equ:conductance_operator} by equation~\ref{equ:conductance_operator_add} and replacing each term by its expectation value, we obtain
\begin{equation}
	\label{equ:conductance_operator_div}
	\left<\hat{C}_{\textrm{nid}}\right>=\frac{\Delta G^{++} - \Delta G^{+-} - \Delta G^{-+} + \Delta G^{--}}{G_{++} + \Delta G^{+-} + \Delta G^{-+} + \Delta G^{--}}\approx \frac{\Delta G^{++} - \Delta G^{+-}}{\Delta G_{++} + \Delta G_{+-}},
\end{equation}
where in the last step we assume $\Delta G^{++} \approx \Delta G^{--}$ and $\Delta G^{+-} \approx \Delta G^{-+}$ because ideally there is no preferred direction. Since in our experiments $\Delta G^{++} - G^{+-} < 0$ and all $\Delta G^{ij}>0$, the spin currents are anti-correlated, as expected for the Cooper pair splitting process.

\subsection{Magnetic Field Sweep Sequence}\hfill

\noindent The determination of the switching fields in extended data figure 9 enables us to define the procedure to obtain the four magnetization states at zero external magnetic field $B = 0$. The measurements in the main text were all done in the following order:
\begin{enumerate}
	\item (-,-): Sweep the external magnetic field to $B = -500\,$mT $<< -B_{\rm sw1}$ in order to form a single magnetic domain along the FSG axis in the negative direction, followed by a sweep back to $B = 0$ to obtain the magnetization state $(-,-)$, since neither of the FSG magnetization switched into the positive orientation.
	\item  (-,+): Continue sweeping from $B=0$ to $B = +45\,$mT $> B_{\rm sw2}$ (but $< B_{\rm sw1}$), followed by a sweep back to $B = 0$ to obtain the magnetization state $(-,+)$.
	\item (+,+): Sweep to $B = +500\,$mT $>> B_{\rm sw1} > B_{\rm sw2}$ to get a single magnetic domain along the $+B$ direction for both FSGs, followed by a sweep back to $B = 0$ to obtain $(+,+)$.
	\item (+,-): Continue sweeping to $B = -45\,$mT $< -B_{\rm sw2}$ (but $> -B_{\rm sw1}$) followed by a sweep back to $B = 0$ to obtain $(+,-)$.
\end{enumerate}
\noindent Similarly, the field sweep sequences used in the experiments for the four magnetization states at $B = \pm 45\,$mT in the main text are as follow (similar for $B = \pm 20\,$mT in extended data figure 7):
\begin{enumerate}
	\item (-,-): Sweep the external magnetic field to $B = -500\,$mT in order to form a single magnetic domain along the FSG axis, followed by a sweep back to $B = -45\,$mT to obtain the magnetization state $(-,-)$ at $B = -45\,$mT.
	\item (-,+): Sweep to $B = +45\,$mT to obtain the magnetization state $(-,+)$ at $B = +45\,$mT.
	\item (+,+): Continue sweeping to $B = +500\,$mT to get a single magnetic domain along the $+B$ direction, followed by a sweep back to $B = +45\,$mT to obtain the $(+,+)$ at $B = +45\,$mT.
	\item (+,-): Continue sweeping to $B = -45\,$mT to obtain $(+,-)$ at $B = -45\,$mT.
\end{enumerate}

\section*{Data Availability Statement}

\noindent All data in the publication are available in numerical form at DOI: \url{https://doi.org/10.5281/zenodo.5805087}.

\bibliography{Reference_CPS_FSG}

\section*{Acknowledgements}

This work has received funding from the Swiss National Science Foundation, the Swiss Nanoscience Institute, the Swiss NCCR QSIT, the FlagERA project, the QuantERA SuperTop project network and the FET Open project AndQC. C.S. has received funding from the European Research Council under the European Union's Horizons 2020 research and innovation programme.
\section*{Author Contributions}

A.Bordoloi fabricated the devices, performed the measurements, analyzed and interpreted the data. V.Z. and L.S. have grown the nanowires. A.Baumgartner helped with the measurements, data analysis and interpretation. A.Bordoloi and A.Baumgartner wrote the paper. C.S. and A.Baumgartner initiated and supervised the project. All authors discussed the results and contributed to the manuscript.   

\section*{Competing Interests}

The authors declare no competing interests.

\end{document}


\maketitle


\newpage

\begin{table} [H]
	\caption{Spin dependence for competing two electron transport processes in a CPS device. We note that LPT + SET (process 4) may mimick the CPS charge signal, but can be distinguished using spin filtering.}
	\label{tab:ProcessesCPS}
	\vspace{0.5cm}
	\centering
	\begin{tabular}{c c c c c c} 
		Process & $\Delta G$ & Parallel state ($P=1$) & Antiparallel state ($P=1$) & $\Delta G^{++}$ & $\Delta G^{+-}$
		\\ \hline \hline \vspace{0.5cm} 
		
		\parbox{2.5cm}{Cooper Pair Splitting (CPS)} & \parbox{3.6cm}{\centering $\propto D_{1}^{\uparrow}D_{2}^{\downarrow} + D_{1}^{\downarrow}D_{2}^{\uparrow}$} & \raisebox{-.3\height}{\includegraphics[width=0.23\textwidth]{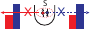}} & \raisebox{-.3\height}{\includegraphics[width=0.23\textwidth]{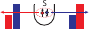}}	& $=0$ & $>0$ \\ \vspace{0.5cm}
		
		\parbox{2.5cm}{Local Pair Tunneling (LPT)} & \parbox{3.6cm}{\begin{align*}
			\Delta G_{1} &\propto  D_{1}^{\uparrow}D_{1}^{\downarrow}\\
			\Delta G_{2} &\propto  D_{2}^{\uparrow}D_{2}^{\downarrow}
		\end{align*}} & 	\raisebox{-.3\height}{\includegraphics[width=0.23\textwidth]{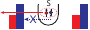}} & \raisebox{-.3\height}{\includegraphics[width=0.23\textwidth]{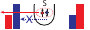}} & $=0$ & $=0$ \\ \vspace{0.5cm}
		
		\parbox{2.5cm}{CPS + Single Electron Tunneling (SET)} & \parbox{3.7cm}{\centering $\propto D_{1}^{\uparrow}D_{2}^{\downarrow}D_{12} + D_{1}^{\downarrow}D_{2}^{\uparrow}D_{12}$} &	\raisebox{-.3\height}{\includegraphics[width=0.23\textwidth]{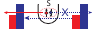}}& \raisebox{-.3\height}{\includegraphics[width=0.23\textwidth]{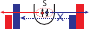}} & $=0$ & $=0$ \\ \vspace{0.5cm}
		
		\parbox{2.5cm}{LPT + SET} & \parbox{3.6cm}{\begin{align*}
			\propto	&D_{1}^{\uparrow}D_{1}^{\downarrow}D_{12} + D_{1}^{\downarrow}D_{1}^{\uparrow}D_{12}\\
				+ &D_{2}^{\uparrow}D_{2}^{\downarrow}D_{12} + D_{2}^{\downarrow}D_{2}^{\uparrow}D_{12}
		\end{align*}} &	\raisebox{-.3\height}{\includegraphics[width=0.23\textwidth]{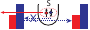}} &	\raisebox{-.3\height}{\includegraphics[width=0.23\textwidth]{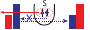}} & $=0$ & $=0$

	\end{tabular}
	
\end{table}

\newpage


\begin{figure}[H]
	\centering
	\includegraphics[width=0.75\columnwidth]{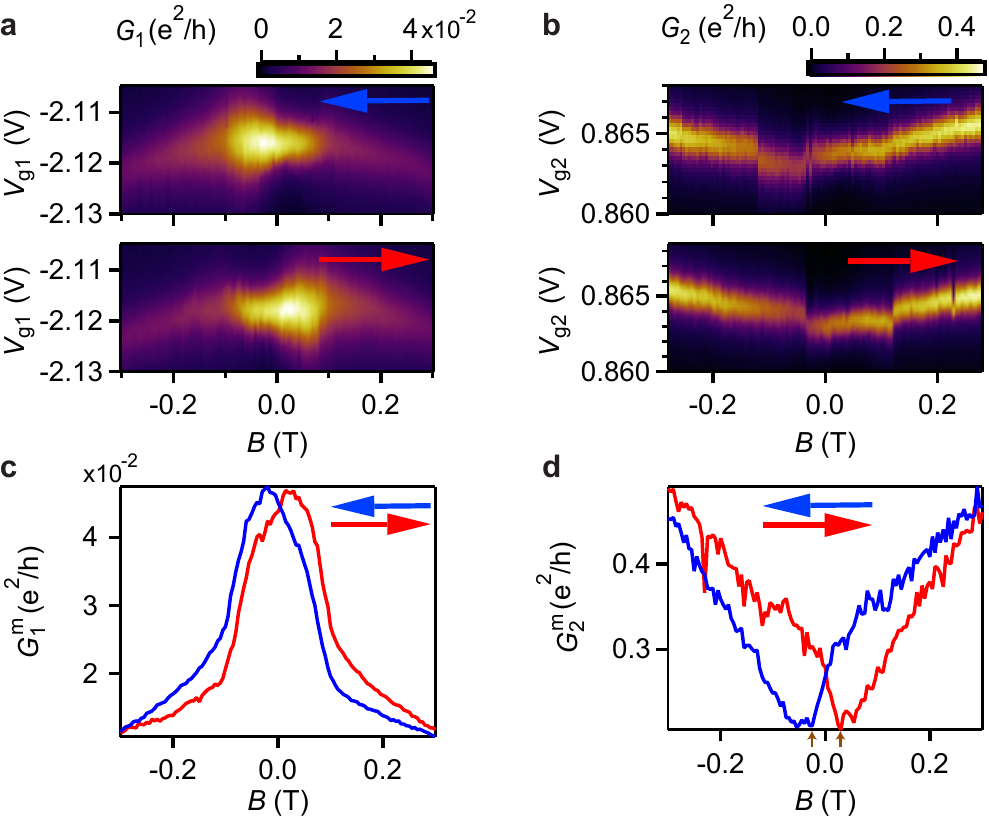}
	\caption{\textbf{Magnetoconductance (MC) measurements.} \textbf{a,b} Down-(blue arrow) and up-sweep (red arrow) maps of $G_{1}$ and $G_{2}$ respectively, as a function of the external magnetic field $B$ and the gate voltage $V_{\textrm{g1}}$ \textbf{(a)} and $V_{\textrm{g2}}$ \textbf{(b)}. \textbf{c,d} Maximum conductance $G_{1}^{\textrm{m}}$ and  $G_{2}^{\textrm{m}}$ as a function of $B$ extracted from \textbf{(a)} and \textbf{(b)} for the up (red) and down (blue) sweep.}
	\label{fig:Fig_S3}
\end{figure} 

In order to demonstrate the presence of a non-zero stray magnetic field $B_{\textrm{str}}$ at each QD position, we measure $G_{1}$ ($G_{2}$) as a function of $V_{\textrm{g1}}$ ($V_{\textrm{g2}}$) for a particular Coulomb blockade resonance of the respective QD at a series of external magnetic fields $B$ applied parallel to the FSG axes. Such maps of $G_{1}$ and $G_{2}$ are shown in extended data figures~\ref{fig:Fig_S3}a and ~\ref{fig:Fig_S3}b for increasing (red arrow) and decreasing (blue arrow) magnetic fields. We note that $B$ is first swept to large values, i.e. $+0.5\,$T or $-0.5\,$T, to ensure the formation of a single domain along the FSG axes. Both, $G_{1}$ and $G_{2}$, show a clear hysteresis with a strong dependence on the sweep direction of $B$, corresponding to a remnant magnetization in both FSGs and a finite $B_{\textrm{str}}$ on both QDs. To explicitly demonstrate this, we extract the maximum conductance $G_{1}^{\textrm{m}}$ and $G_{2}^{\textrm{m}}$ for QD1 and QD2, as plotted in extended data figures~\ref{fig:Fig_S3}c and ~\ref{fig:Fig_S3}d respectively, for increasing (red) and decreasing (blue) $B$.

In the up sweep, we find that $G_{2}^{\textrm{m}}$ decreases roughly linearly with increasing $B$ followed by a relatively sharp decrease around $B=0$. At $B \approx 25\,$mT, $G_{2}^{\textrm{m}}$ achieves it minimum value, and starts to increase with more positive $B$. The down sweep can be described in a similar manner, but mirror symmetric around $B=0$. Qualitatively, $G_{2}^{\textrm{m}}$ can be understood as a smooth magnetoconductance (MC) of QD2 \cite{Fabian2016,Bordoloi2020}, which changes relatively abruptly when the corresponding FSG reverses its magnetization direction. The red curve shows a relatively abrupt change at $B = B_{\textrm{sw2}} \approx 25\,$mT at which the FSG magnetization is reoriented, as indicated by the brown arrows in extended data figure~\ref{fig:Fig_S3}d.

Similar to $G_{2}^{\textrm{m}}$, $G_{1}^{\textrm{m}}$ exhibits a clear hysteresis around $B=0$ as shown in extended data figure~\ref{fig:Fig_S3}c. In the up-sweep (red curve), $G_{1}^{\textrm{m}}$ increases roughly linearly with increasing $B$ up to $B = -110\,$mT, followed by a steep increase with a maximum at $B = 20\,$mT and a steep decrease around $B \approx 100-105\,$mT. For $B > 100-105\,$mT, $G_{1}^{\textrm{m}}$ shows a roughly linear decrease with more positive $B$. The down sweep (blue curve) is mirror-symmetric with respect to the up sweep (red curve) at $B=0$. 

\newpage


\begin{figure}[H]
	\centering
	\includegraphics[width=0.75\columnwidth]{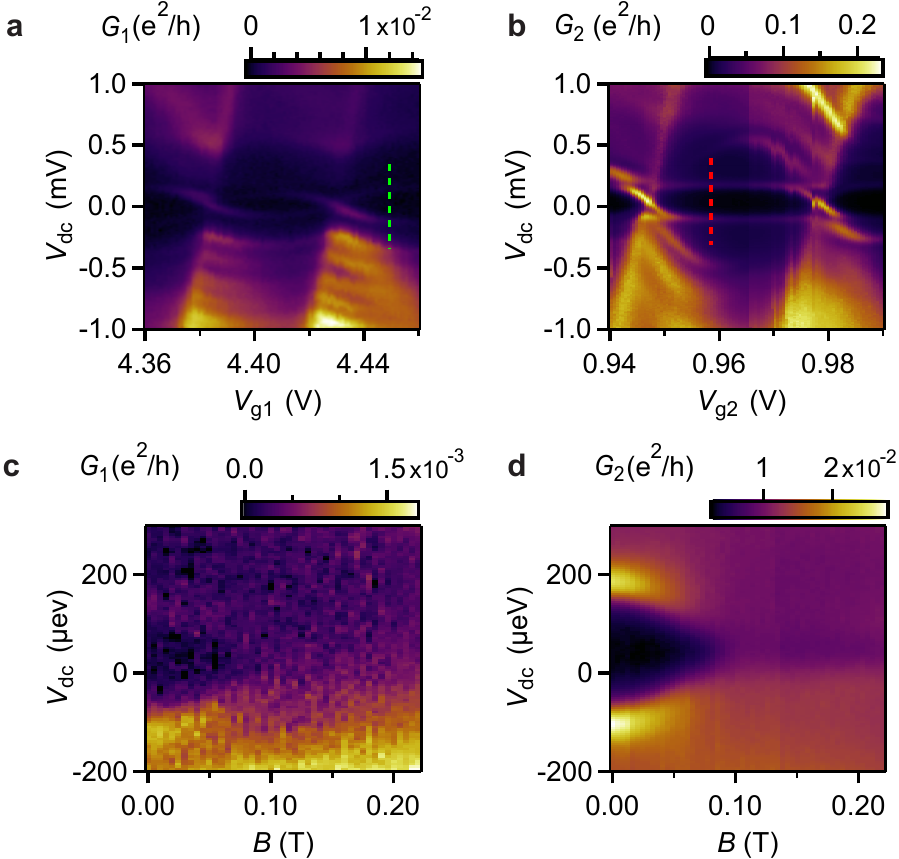}
	\caption{\textbf{Bias spectroscopy and superconducting gap measurement.} \textbf{a,b} Differential conductances $G_{1}(V_{\rm g1})$ and $G_{2}(V_{\rm g2})$ respectively, as a function of the source-drain bias voltage $V_{\textrm{dc}}$. \textbf{c,d} $G_{1}$ and $G_{2}$ as a function of $V_{\textrm{dc}}$ and the external magnetic field $B$ for the cross sections marked by the green and red dashed lines in \textbf{(a)} and \textbf{(b)} respectively. Both measurements show a critical magnetic field of $B_{\textrm{c}} \approx 90\,$mT for the superconducting contact.}
	\label{fig:Fig_S8}
\end{figure}

In this section, we present bias spectroscopy experiments for the individual QDs to determine their transport properties at zero external magnetic field, $B = 0$, applied in the substrate plane along the FSG long axes. Extended data figures~\ref{fig:Fig_S8}a and ~\ref{fig:Fig_S8}b show colorscale plots of $G_{1}$ and $G_{2}$ as a function of the applied bias voltage $V_{\textrm{dc}}$ and the gate voltages $V_{\textrm{g1}}$ and $V_{\textrm{g2}}$, respectively. We extract an addition energy of $E_{\textrm{add,1}} \approx 1.6\,$meV and $E_{\textrm{add,2}} \approx 2.5\,$meV for QD1 and QD2, and a superconducting gap of $\Delta \sim 110\,\mu$eV in both datasets. The cross lever arms (across the the S contact) are one order of magnitude lower compared to the lever arm of a close by FSG, which allows one to independently tune the QDs. In order to characterize the superconductor, we apply an in-plane external magnetic field $B$ parallel to the FSG long axes. $G_{1}$ and $G_{2}$ as a function of $V_{\textrm{dc}}$ and $B$, along the cross sections marked by the green and red dashed line in extended data figures~\ref{fig:Fig_S8}a and ~\ref{fig:Fig_S8}b, are plotted in extended data figures~\ref{fig:Fig_S8}c and ~\ref{fig:Fig_S8}d, respectively. We find that the superconducting gap is suppressed at a critical magnetic field of $B_{\textrm{c}} \approx 90\,$mT. Clearly, the magnitudes of $B_{\textrm{str1}}$ and $B_{\textrm{str2}}$ cannot be larger than $B_{\textrm{c}}$ at the position of the superconductor, and are much smaller than can be resolved here, considering that $\Delta^{*}(B)$ shows a maximum ($\frac{d\Delta^{*}}{dB} = 0$) at $B\approx0$, i.e. no detectable offset field at the position of S.

\newpage


\begin{figure}[H]
	\centering
	\includegraphics[width=0.5\columnwidth]{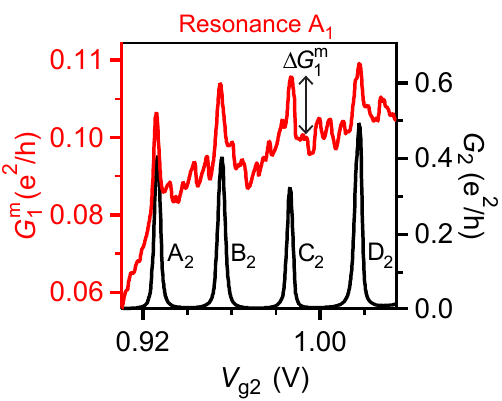}
	\caption{\textbf{$\Delta G_{1}^{\textrm{m}}$ for the QD resonance A$_{1}$ in figure 1d of the main text .} Maximum conductance $G^{\textrm{m}}_{1}$ as a function of gate voltage $V_{\textrm{g2}}$ for resonance $\textrm{A}_{1}$ of QD1, showing peaks whenever QD2 is tuned across any of the Coulomb blockade resonances A$_{2}$-D$_{2}$.}
	\label{fig:Fig_S10}
\end{figure}

\newpage


\begin{figure}[H]
	\centering
	\includegraphics[width=0.8\columnwidth]{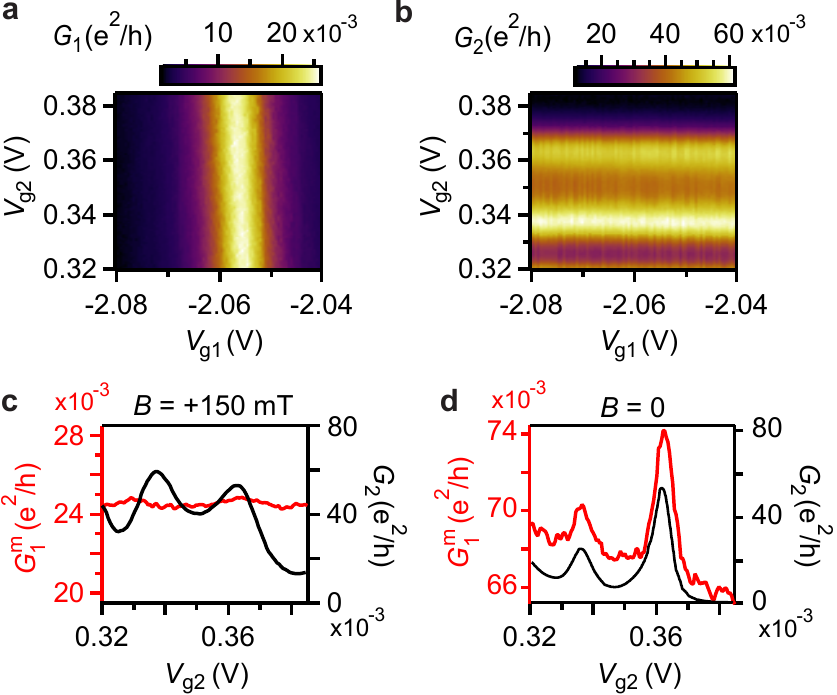}
	\caption{\textbf{Normal-state measurements at $B=+150\,$mT.} \textbf{a,b} Differential conductances $G_{1}$ and $G_{2}$ respectively, measured simultaneously as a function of $V_{\textrm{g1}}$ and $V_{\textrm{g2}}$ at a bias voltage of $V_{\textrm{dc}} = 0$ and an external magnetic field of $B=+150\,$mT. \textbf{c} Maximum conductance $G^{\textrm{m}}_{1}$ as a function of gate voltage $V_{\textrm{g2}}$ at $B=+150\,$mT (red curve), showing a much smaller modulation and no obvious correlation to $G_{2}$, when the latter (black curve) is tuned through Coulomb blockade peaks using $V_{\textrm{g2}}$. \textbf{d} Maximum conductance $G^{\textrm{m}}_{1}$ as a function of gate voltage $V_{\textrm{g2}}$ at $B=0$ for the same resonances as in \textbf{c}, showing peaks when $G_{2}$ is tuned across Coulomb blockade resonances by $V_{\textrm{g2}}$. We note that the scale of $G^{\textrm{m}}_{1}$ is adjusted to show the same conductance span.}
	\label{fig:Fig_S5}
\end{figure}

\newpage


\begin{figure}[H]
	\centering
	\includegraphics[width=0.9\columnwidth]{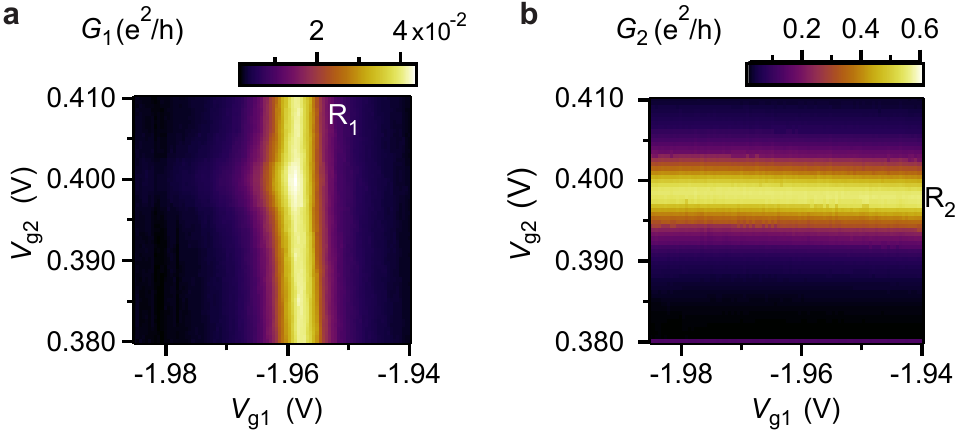}
	\caption{\textbf{Data supporting figure 2 of the main text.} \textbf{a,b} Differential conductances $G_{1}$ and $G_{2}$ respectively, as a function of $V_{\textrm{g1}}$ and $V_{\textrm{g2}}$ at zero bias voltage, $V_{\textrm{dc}} = 0$ and zero external magnetic field, $B=0$, for the resonance crossing (R$_1$, R$_2$) described in figure 2 in the main text.}
	\label{fig:Fig_S9}
\end{figure}

\newpage


\begin{figure}[H]
	\centering
	\includegraphics[width=0.85\columnwidth]{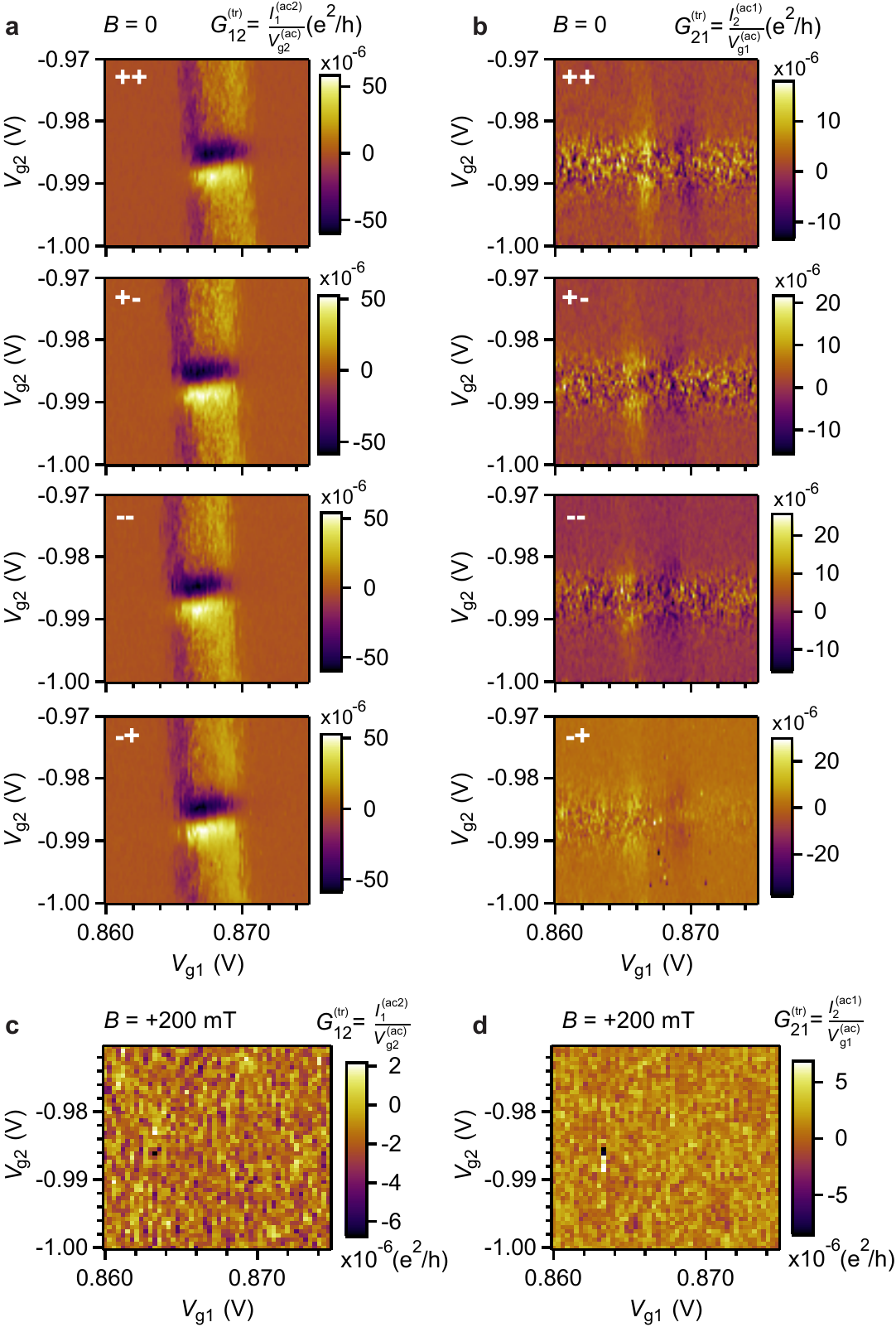}
	\caption{\textbf{Transconductance Measurements for the four magnetization states at $B=0$ and $B=+200\,$mT.} \textbf{a,b,c,d} Transconductance $G^{\textrm{(tr)}}_{12} = \frac{I_{1}^{\textrm{(ac2)}}}{V_{\textrm{g2}}^{\textrm{(ac)}}}$ (\textbf{a,c}) and $G^{\textrm{(tr)}}_{21} = \frac{I_{2}^{\textrm{(ac1)}}}{V_{\textrm{g1}}^{\textrm{(ac)}}}$ (\textbf{b,d}) measured as a function of $V_{\textrm{g1}}$ and $V_{\textrm{g2}}$ with a bias of $V_{\textrm{dc}} = 25\mu$eV applied to S, for each magnetization state ($j,k$) indicated in each figure, at $B=0$ (\textbf{a,b}) and $B=+200\,$mT (\textbf{c,d}) for the resonance crossings M$_{1}$ and N$_{2}$ in figure~3 of the main text. We do not observe any modulation of the transconductance if S is in the normal-state ($B=+200\,$mT).}
	\label{fig:Fig_S6}
\end{figure}

 \newpage 


\begin{figure}[H]
	\centering
	\includegraphics[width=\columnwidth]{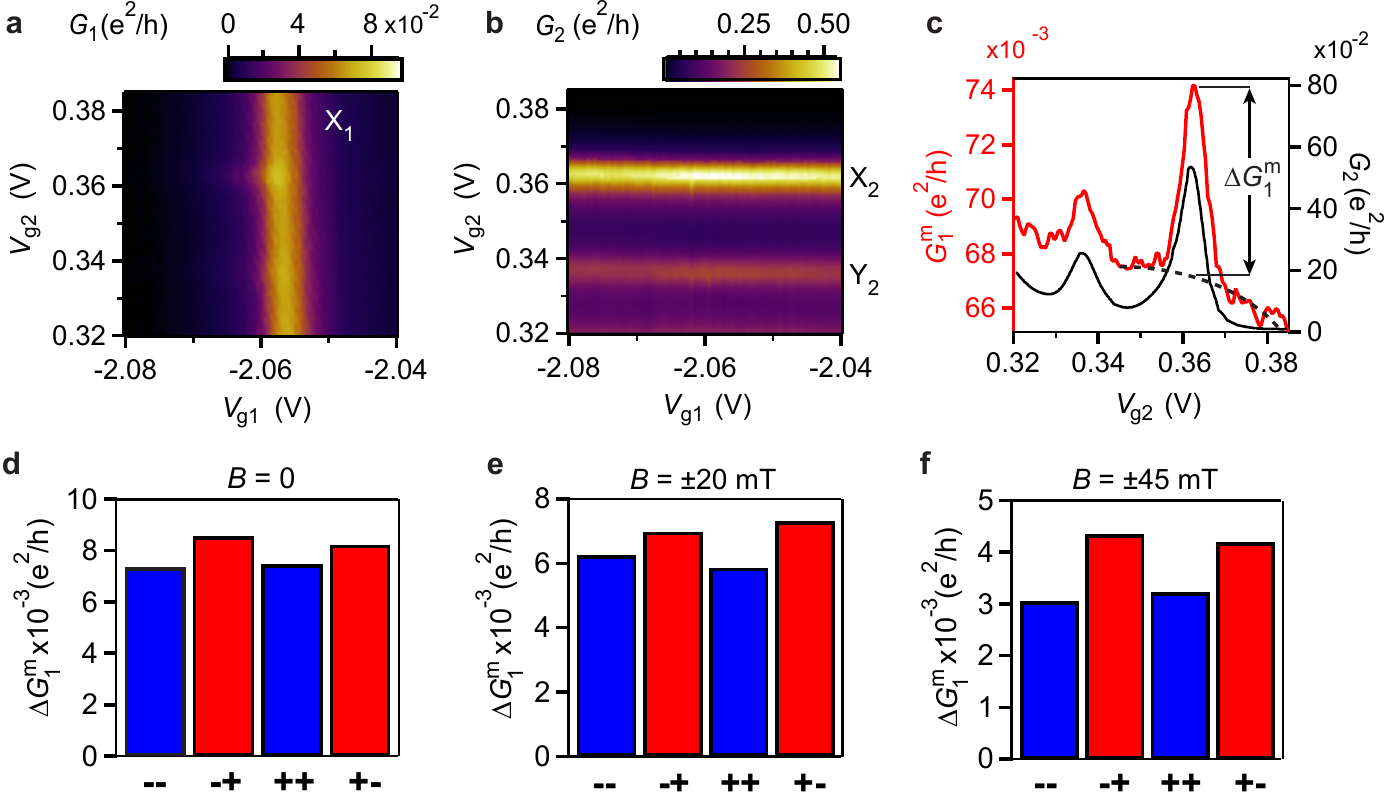}
	\caption{\textbf{Conductance maxima modulation at finite magnetic fields for main text figure 4.} \textbf{a,b} Differential conductances $G_{1}$ and $G_{2}$ respectively, measured as a function of $V_{\textrm{g1}}$ and $V_{\textrm{g2}}$ at zero bias $V_{\textrm{dc}} = 0$ and $B=0$. \textbf{c} Maximum conductance $G^{\textrm{m}}_{1}$ as a function of the gate voltage $V_{\textrm{g2}}$ for the resonances in \textbf{a} and \textbf{b}, showing peaks when $G_{2}$ is tuned across Coulomb blockade resonances by $V_{\textrm{g2}}$. \textbf{d,e,f} Modulation of the conductance maximum, $\Delta G^{\textrm{m}}_{1}$, for all four magnetization states ($j,k$) measured at $B=0$ \textbf{(d)}, $B=\pm20\,$mT \textbf{(e)}, and $B=\pm45\,$mT \textbf{(f)} for the resonance crossing (X$_1$,X$_2$).}
	\label{fig:Fig_S7}
\end{figure}

\newpage


\begin{figure}[H]
	\centering
	\includegraphics[width=0.5\columnwidth]{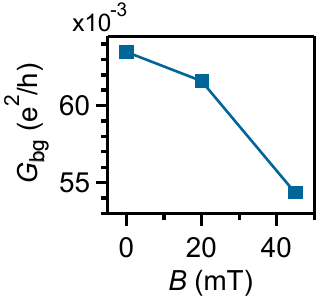}
	\caption{\textbf{Background conductance $G_{\textrm{bg}}$ versus the external magnetic field $B$.} Here we plot the background conductance at the respective resonance position, extracted from the parabolic fits discussed in the main text. This background is most probably dominated by LPT processes. We note that $G_{\textrm{bg}}$ decreases with increasing $B$, as expected for the local processes (see extended data table~\ref{tab:ProcessesCPS}) and for an increasing QD spin polarization.}
	\label{fig:Fig_S11}
\end{figure}

\newpage


\begin{figure}[H]
	\centering
	\includegraphics[width=\columnwidth]{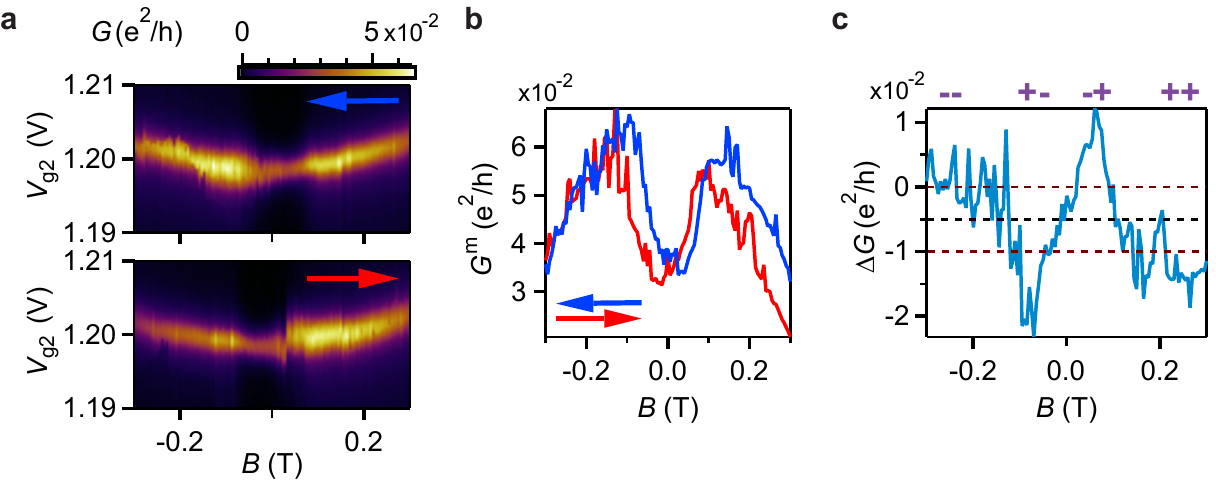}
	\caption{\textbf{Estimates for the characteristic switching fields B$_{\textrm{sw}}$.} \textbf{a} Down-(blue arrow) and up-sweep (red arrow) maps of the differential conductance $G$ as a function of $B$ and the gate voltage $V_{\textrm{g2}}$ for a two-terminal measurement between the two normal (N) contacts, with the middle superconducting contact (S) floating. \textbf{b} Maximum conductance $G^{\textrm{m}}$ as a function of $B$ extracted from figure~\textbf{a} for the up (red) and down (blue) sweep. \textbf{c} $\Delta G = G_{\textrm{up}}^{\textrm{m}} - G_{\textrm{down}}^{\textrm{m}}$ as a function of $B$ derived from figure~\textbf{b}, showing estimates for all parallel and antiparallel magnetization states.}
	\label{fig:Fig_S2}
\end{figure}

To estimate the characteristic switching fields of the FSGs, we operate the device in a spin-valve configuration, i.e. we perform standard two-terminal lock-in measurements with the two QDs in series and with S kept floating, and measure the differential conductance $G$ as a function of $B$ and $V_{\textrm{g2}}$ for increasing (red) and decreasing (blue) $B$ as shown in extended data figure~\ref{fig:Fig_S2}a. The corresponding maximum conductance $G^{\textrm{m}}$ is plotted in extended data figure~\ref{fig:Fig_S2}b. In the up sweep (red curve), $G^{\textrm{m}}$ first increases with increasing $B$, followed by a maximum at $B \approx -130\,$mT and a subsequent decrease with a minimum at $B \approx -20\,$mT. Around $B \approx 0$,  $G^{\textrm{m}}$ starts to increase again with small positive $B$, followed by another maximum at $B \approx 95\,$mT, and a further decrease towards more positive $B$. The down sweep (blue curve) is mirror symmetric to the up sweep (red curve) at $B=0$. 

To determine the parallel and antiparallel magnetization states, we plot $\Delta G = G_{\textrm{up}}^{\textrm{m}} - G_{\textrm{down}}^{\textrm{m}}$, i.e. we subtract the blue from the red curve, as a function of $B$ in extended data figure~\ref{fig:Fig_S2}c, where $G_{\textrm{up}}^{\textrm{m}}$ and $G_{\textrm{down}}^{\textrm{m}}$ refer to the up (red) and down (blue) sweep in extended data figure~\ref{fig:Fig_S2}b. We first assign an effective average zero level of the measured data, as indicated by the black dashed line in extended data figure~\ref{fig:Fig_S2}c. We then define a lower and upper conductance limit (brown dashed lines) for significant deviation of $\Delta G$ from the average zero. We use the $B$ values at which the upper horizontal brown dashed line meets $\Delta G$ as the two switching fields, $B_{\textrm{sw1}} \approx 100\,$mT and $B_{\textrm{sw2}} \approx 25\,$mT, respectively. We obtain similar switching field values for the lower horizontal brown dashed line. Therefore, for magnetic field $B$ values between $ B_{\textrm{sw1}} > B > B_{\textrm{sw2}}$, the FSGs are oriented in an antiparallel magnetization configuration.



\bibliographystyle{nature}
\renewcommand{\refname}{References}